\documentclass[journal=jctcce,manuscript=article,layout=traditional]{achemso}

\pdftrailerid{}
\usepackage[T1]{fontenc} \usepackage{mathtools}
\usepackage{graphicx}
\usepackage{lineno}
\usepackage{amsmath}
\usepackage{amssymb}
\usepackage{physics}
\usepackage{mathrsfs}
\usepackage{bm}
\usepackage{threeparttable}
\usepackage{float}
\usepackage{longtable}
\usepackage{multirow}
\usepackage{booktabs}
\usepackage{hyperref}
\usepackage[section]{placeins}
\usepackage{makecell}
\usepackage{xcolor}

\newcommand{\recZn}{37.95 \pm 0.77}
\newcommand{\recCd}{45.68 \pm 1.21}
\newcommand{\recHg}{34.04 \pm 0.68}
\newcommand{\recCn}{27.92 \pm 0.28}

\SectionNumbersOn

\makeatletter
\newcommand{\warninput}[1]{\filename@parse{#1}\InputIfFileExists{#1}{}{\message{LaTeX Warning: File `\filename@base.\ifx\filename@ext\relax tex\else\filename@ext\fi' not found on input line \the\inputlineno}}}
\makeatother

\title{Relativistic and electron-correlation effects in static dipole polarizabilities for group 12 elements}

\author{YingXing Cheng}
\affiliation[University of Stuttgart]{Institute of Applied Analysis and Numerical Simulation, University of Stuttgart, Pfaffenwaldring 57, 70569, Stuttgart, Germany}
\email{yingxing.cheng@mathematik.uni-stuttgart.de}

\begin{document}

    \begin{abstract}
        In this study, we report a comprehensive calculation of the static dipole polarizabilities of group 12 elements using the finite-field approach combined with the relativistic coupled-cluster method, including single, double, and perturbative triple excitations.
Relativistic effects are systematically investigated, including scalar-relativistic, spin-orbit coupling (SOC), and fully relativistic Dirac-Coulomb contributions.
The final recommended polarizability values are $37.95 \pm 0.77$ a.u. for Zn, $45.68 \pm 1.21$ a.u. for Cd, $34.04 \pm 0.68$ a.u. for Hg, and $27.92 \pm 0.28$ a.u. for Cn.
These results are in excellent agreement with the 2018 Table of static dipole polarizabilities for neutral atoms [Mol. Phys. \textbf{117}, 1200 (2019)] and provide reduced uncertainties for Cd and Cn.
Our analysis shows that scalar-relativistic effects dominate the relativistic corrections, with SOC contributions found to be negligible.
The role of electron correlation is thoroughly examined across the non-relativistic, scalar-relativistic, and fully relativistic Dirac-Coulomb regimes, underscoring its critical importance in achieving accurate polarizability predictions.
     \end{abstract}


    \newpage

    \section{Introduction}
    \label{sec:introduction}
    The electric dipole polarizability characterizes the deformation of electron density of a system in response to an external electric field and plays a crucial role in understanding interactions in atomic and molecular physics, such as scattering cross sections, refractive indices, dielectric constants, and interatomic forces.\cite{Schwerdtfeger2019}
It also serves as a benchmark for theoretical methods like density functional theory\cite{Bast2008} and aids in developing basis sets for computational chemistry.\cite{Dyall2011,Dyall2016,Dyall2023,Ferreira2020,CanalNeto2021,Neto2021,Centoducatte2022,Neto2023,Gomes2024,Sampaio2024}
Furthermore, accurate dipole polarizabilities enhance the precision of optical atomic clocks by mitigating black-body radiation shifts.\cite{Ludlow2015,Safronova2012a}

Schwerdtfeger and Nagle recently compiled a comprehensive table of accurate dipole polarizabilities for neutral atoms with nuclear charges $Z=1$ to $120$ (excluding livermorium, $Z=116$).\cite{Schwerdtfeger2019}
The latest version is available in Ref.~\citenum{Schwerdtfeger2023}.
While precise values exist for lighter elements, such as helium ($Z=2$)\cite{Schmidt2007} and neon ($Z=10$),\cite{Gaiser2010} accurate polarizabilities for heavier elements remain scarce.\cite{Schwerdtfeger2019}
Experimental determination of dipole polarizabilities for heavier elements is challenging, necessitating the use of computational methods.
Accurate predictions require precise treatment of electron correlation within a relativistic framework, particularly for heavy atoms.
The coupled-cluster (CC) method with single and double excitations and perturbative triples, CCSD(T), is widely regarded as the gold standard for these calculations.
However, relativistic CCSD(T) implementations for open-shell atoms with multiple open electrons are underdeveloped.
In such cases, the relativistic multi-reference configuration interaction (MRCI) method has been employed.\cite{Olsen1990,Fleig2003,Fleig2006,Knecht2010,Fleig2012,Schwerdtfeger2019}
Cheng recently calculated dipole polarizabilities for main-group elements (excluding hydrogen) using fully relativistic CCSD(T) and MRCI methods with extensive Dyall\cite{Dyall2002,Dyall2004,Dyall2006,Dyall2007,Dyall2009,Dyall2010,Dyall2011,Dyall2016} and ANO-RCC basis sets.\cite{Roos2004,Roos2005}
This procedure was later extended to group 11 elements, including Cu ($Z=29$), Ag ($Z=47$), and Au ($Z=79$), using the relativistic CCSD(T) method.\cite{Cheng2024a}

For group 12 elements, CC methods benefit from the closed-shell electronic structure of these systems.
Two primary approaches, analytical and finite-difference methods, are used for calculating properties like dipole polarizabilities.
Analytical methods derive properties directly from expectation values of the wave function and can be categorized into sum-over-states and linear-response coupled-cluster (LRCC) methods.
The sum-over-states method depends on the inclusion of intermediate states, including continuum contributions, which is computationally demanding.
Alternatively, LRCC solves inhomogeneous equations for property calculations,\cite{Monkhorst1977} but this approach becomes prohibitively expensive for heavy elements in a relativistic framework.\cite{Yuan2024}
Perturbed relativistic coupled-cluster (PRCC) methods offer another option by applying perturbation theory to compute first-order perturbed wave functions.
The PRCC method has been used to calculate dipole polarizabilities for Zn ($Z=30$),\cite{Singh2014,Chattopadhyay2015} Cd ($Z=48$),\cite{Singh2014,Chattopadhyay2015,Sahoo2018b} Hg ($Z=80$),\cite{Singh2015,Chattopadhyay2015,Sahoo2018,Kumar2021} and Cn ($Z=112$).\cite{Kumar2021}
However, PRCC may yield inconsistent results in some cases, such as for Hg.\cite{Sahoo2018d}
The relativistic normal coupled-cluster (RNCC) method has been developed as an alternative, offering improved accuracy by naturally terminating expectation values and satisfying the Hellmann-Feynman theorem.\cite{Bishop1991,Kowalski2004,Hagen2014,Sahoo2018d}

Finite-difference methods provide a simpler alternative by numerically evaluating second derivatives of energy with respect to the electric field.
When combined with least-squares fitting, these methods avoid numerical instabilities often associated with higher-order derivatives.\cite{Kassimi1994}
This approach has been applied to group 12 elements, including Zn,\cite{Kellö1995,Goebel1996,Roos2005} Cd,\cite{Kellö1995,Goebel1996,Seth1997,Guo2021,Zaremba-Kopczyk2021,Roos2005} Hg,\cite{Kellö1995,Seth1997,Pershina2008b,Borschevsky2015} and Cn.\cite{Seth1997,Pershina2008b}
However, studies combining finite-difference methods with least-squares fitting remain limited.
To the best of our knowledge, no comprehensive study has applied the CC method within a fully relativistic Dirac-Coulomb framework to compute dipole polarizabilities of group 12 elements under nonrelativistic, scalar-relativistic, and fully relativistic conditions.
This work aims to fill this gap.

In this study, we employ the relativistic CCSD(T) method, combined with finite-difference and least-squares procedures, to calculate atomic dipole polarizabilities for group 12 elements.
Our results agree with the recommended values in Ref.~\citenum{Schwerdtfeger2019} and align with other theoretical predictions in the literature.
Additionally, we provide a detailed analysis of relativistic and electron-correlation effects, based on Refs.~\citenum{Cheng2024a,Cheng2024b}.
The final recommended values are determined using an error-estimation method proposed in these references.

The remainder of this paper is organized as follows.
Section~\ref{sec:methods} describes the computational methods, while Sec.~\ref{sec:details} provides the computational details.
The results are presented and discussed in Sec.~\ref{sec:results}, and a summary of the findings is provided in Sec.~\ref{sec:summary}.
Throughout this work, atomic units are used unless otherwise specified.

    \section{Methods}
    \label{sec:methods}
    The Dirac-Coulomb (DC) Hamiltonian is defined as\cite{Fleig2012}
\begin{align}
    \hat{H}_\text{DC} = \sum_i \hat{h}_\text{D}(i) + \sum_{i<j} \frac{1}{r_{ij}} + \sum_{A<B} V_{AB},
    \label{eq:H_DC}
\end{align}
where $\hat{h}_\text{D}$ represents the one-electron Dirac Hamiltonian, $V_{AB}$ denotes the nuclear-nuclear interaction, and $1/r_{ij}$ corresponds to the Coulombic electron-electron interaction.
Breit and quantum electrodynamics (QED) effects are considered only in the uncertainty estimation.

Due to the high computational cost of four-component relativistic calculations, two-component approximations such as the Douglas-Kroll-Hess (DKH) Hamiltonian~\cite{Douglas1974,Hess1985,Hess1986} and the zeroth-order regular approximation (ZORA)~\cite{Chang1986,vanLenthe1994,vanLenthe1996} are commonly employed.
These methods, however, are limited to finite-order corrections and may not fully account for all relativistic effects.
To overcome these limitations, the exact two-component (X2C) method~\cite{Ilias2007} provides an infinite-order relativistic treatment.
This method decomposes relativistic effects into scalar-relativistic (spin-free) and spin-orbit coupling (SOC) contributions,\cite{Dyall2001} thereby achieving higher accuracy without the computational expense of four-component methods.
Readers are directed to Refs.~\citenum{Dyall1997,Dyall2002a,Kutzelnigg2005,Filatov2007,Peng2007,Ilias2007,Liu2009,Saue2011,Peng2012,Liu2014,Liu2016a,Liu2020} for further details on the X2C method.

The X2C molecular mean-field (X2Cmm) approach~\cite{Sikkema2009} further eliminates the integral transformation of relativistic atomic orbital two-electron integrals, which significantly reduces the cost of the integral transformation in CC calculations.
This reduction in computational overhead makes the method particularly effective for medium-sized molecular systems, as demonstrated in CC calculations.\cite{Zhang2024}

In this study, the X2Cmm method is primarily used for fully relativistic DC calculations.
For nonrelativistic and scalar-relativistic properties, the spin-free X2C method is employed,\cite{Dyall2001} as described in Refs.~\citenum{Saue2011,Saue2020}.

Uncorrelated reference calculations are performed using the Dirac-Hartree-Fock (DHF) method.
The terms ``orbital'' and ``spinor'' are used to describe nonrelativistic and relativistic electronic states, respectively.
Correlation effects are included through second-order M\o ller-Plesset perturbation theory (MP2)~\cite{vanStralen2005a} and coupled-cluster methods, including the CC method with single and double excitations (CCSD) and CCSD(T).\cite{Visscher1996}

The abbreviations NR-CC, SR-CC, and DC-CC refer to coupled-cluster calculations performed under nonrelativistic, scalar-relativistic, and fully relativistic DC approximations, respectively.
To manage computational cost, orbital or spinor spaces are divided into inner-core, outer-core, valence, and virtual regions, with correlation applied only to the outer-core and valence spaces.

Dipole polarizabilities, $\alpha_m^n$, are computed using a chosen correlation method $m$ (DHF, MP2, CCSD, or CCSD(T)) and relativistic treatment $n$ (NR, SR, or DC).
The correlation contribution to polarizability with a specific relativistic effect $n$ is defined as
\begin{align}
    \Delta \alpha_c^n := \alpha_{\text{CCSD(T)}}^n - \alpha_{\text{DHF}}^n.
    \label{eq:alpha_corr}
\end{align}
Scalar-relativistic and SOC contributions are quantified as
\begin{align}
    \Delta \alpha_r^\text{SR} &:= \alpha_\text{CCSD(T)}^\text{SR} - \alpha_\text{CCSD(T)}^\text{NR}, \\
    \Delta \alpha_r^\text{SOC} &:= \alpha_\text{CCSD(T)}^\text{DC} - \alpha_\text{CCSD(T)}^\text{SR}.
    \label{eq:rel_effects_sr_soc}
\end{align}
The total relativistic correction is the sum of these contributions: $\Delta \alpha_r^\text{DC} = \Delta \alpha_r^\text{SR} + \Delta \alpha_r^\text{SOC}$.

Static dipole polarizabilities are determined using the finite-field approach.\cite{Das1998}
The atomic energy in an external field $F_z$ applied along the $z$-axis is expanded as
\begin{align}
    E(F_z) \approx E_0 - \frac{1}{2}\alpha F_z^2 - \frac{1}{24}\gamma F_z^4,
    \label{eq:ff_d_2}
\end{align}
where $E_0$ is the field-free energy, $\alpha$ represents the polarizability, and $\gamma$ is the hyperpolarizability.
Least-squares fitting is applied to extract $\alpha$ and $\gamma$ from calculated energies.\cite{pydirac}
In cases where $\gamma$ yields unphysical results,\cite{Cheng2024b} a simplified model retaining only $\alpha$ is adopted:
\begin{align}
    E(F_z) \approx E_0 - \frac{1}{2}\alpha F_z^2.
    \label{eq:ff_d_1}
\end{align}

To address unphysical $\gamma$ values more appropriately, an approximate, valid positive value for $\gamma$, denoted as $\gamma_\text{approx.}$, is assigned.
In this case, Eq.~\eqref{eq:ff_d_2} becomes
\begin{align}
    E(F_z) \approx E_0 - \frac{1}{2}\alpha F_z^2 - \frac{1}{24} \gamma_\text{approx.} F_z^4.
    \label{eq:ff_d_1_fixed}
\end{align}

Uncertainties in polarizabilities are evaluated using a composite scheme,\cite{Kallay2011,Yu2015,Irikura2021,Cheng2024a} with the total uncertainty expressed as
\begin{align}
    P_\text{final} = P_\text{CCSD} + \Delta P_\text{basis} + \Delta P_\text{(T)} + \Delta P_\text{core} + \Delta P_\text{vir} + \Delta P_\text{fitting} + \Delta P_\text{SOC} + \Delta P_\text{others},
\end{align}
where the terms correspond to contributions from basis set size, triple excitations, core correlation, virtual truncation, fitting errors, SOC effects, and other sources.
Interested readers are referred to Ref.~\citenum{Cheng2024a} for further details.
Contributions from Breit and QED interactions are included in the ``others'' category based on literature data.\cite{Dutta2020,Kumar2021}

    \section{Computational Details}
    \label{sec:details}
    In this study, we employ uncontracted Dyall quadruple-$\zeta$ basis sets\cite{Dyall2004,Dyall2007,Dyall2010} along with the ANO-RCC basis set.\cite{Roos2004,Roos2005}
The original dyall.cv4z basis sets are augmented with additional even-tempered functions to extend each function type.
The exponential coefficients of the added functions are determined by the relation $\zeta_{N+1} = \zeta_N^2 /\zeta_{N-1}$, where $\zeta_N$ and $\zeta_{N-1}$ are the smallest exponents in each atomic shell of the default basis.\cite{Yu2015}
The augmented basis sets are labeled as s-aug-dyall.cv4z for single augmentations and d-aug-dyall.cv4z for double augmentations.
Similarly, single augmentations of the ANO-RCC basis set are labeled as s-aug-ANO-RCC.

Orbitals within an energy range of $-20$ to $25$ a.u.\ are included in the correlation space for our CC calculations, with a convergence threshold set to $10^{-10}$.
Electric fields with strengths of $0.000$, $0.0005$, $0.001$, $0.002$, and $0.005$ a.u.\ are applied to each system to compute dipole polarizabilities.
All calculations are performed using the \texttt{DIRAC18} package.\cite{DIRAC18}
The computed energies are fitted to Eqs.~\eqref{eq:ff_d_2}-\eqref{eq:ff_d_1_fixed} using a least-squares method implemented as described in Ref.~\citenum{pydirac} to derive the dipole polarizabilities.

In the CC calculations, various correlation methods, including DHF, MP2, CCSD, and CCSD(T), are employed.\cite{DIRAC18}
Each CC calculation is identified by a unique string descriptor that specifies the computational method, basis set, and correlation level.
For example, ``1C-SR-CC@s-aug-dyall.cv4z@(core 3)[vir 279]'' consists of three components separated by the delimiter ``@''.
The first component denotes the computational method, such as NR-CC, SR-CC, or DC-CC, with ``1C'' or ``2C'' indicating a one-component or two-component relativistic Hamiltonian, respectively.
For NR-CC and SR-CC, the SOC effect is not included; thus, the Hamiltonian in all calculations is one-component.
It should be noted that in Refs.~\citenum{Cheng2024a,Cheng2024b}, ``2C'' is used for NR-CC and SR-CC, where ``2C'' refers to the X2C module being active.
All DC-CC calculations use a two-component Hamiltonian in this work.

The second component specifies the basis set, e.g., s-aug-dyall.cv4z.
The final component describes the correlation level, specifying the number of active electrons and virtual orbitals in the format ``(core $N$)[vir $M$]'', where $N$ is the number of outer-core and valence electrons, and $M$ represents the number of virtual orbitals.

The percentage error $\delta_m$ of a property $X = \alpha~\text{or}~\gamma$ is defined as
\begin{align}
    \delta_m = \frac{X_m - X_\text{CCSD(T)}}{X_\text{CCSD(T)}} \times 100\%,
\end{align}
where $m$ represents one of the methods (DHF, MP2, or CCSD), and $X_\text{CCSD(T)}$ denotes the results obtained using CCSD(T), which are taken as the reference.

In this work, we consider only the atomic dipole polarizabilities for the ground state of group 12 elements, i.e., the $^1S$ and $^1S_0$ states in $LS$ and $jj$ coupling, respectively.
Here, $LS$ coupling is used for all NR-CC and SR-CC results, whereas $jj$ coupling is applied to all DC-CC values.
The core occupations for Zn, Cd, Hg, and Cn are $3d^{10}$, $4d^{10}$, $5d^{10}$, and $6d^{10}$, respectively.

    \section{Results}
    \label{sec:results}
    \subsection{Zn}

Table~\ref{tab:dipole_group__12__Zn__2__0} presents the dipole polarizabilities $\alpha$ of Zn, determined by fitting Eq.~\eqref{eq:ff_d_2}.
The corresponding hyperpolarizabilities $\gamma$ are listed in Table S1 in the Supporting Information.
The values of $\alpha$ at the SR CCSD(T)/s-aug-dyall.cv4z level show convergence compared to those obtained with dyall.cv4z and d-aug-dyall.cv4z.
Thus, the s-aug-dyall.cv4z basis set is used for the most accurate calculations at each relativistic level.

\begin{table}[ht]
\centering

            \caption{
            Dipole polarizabilities ($\alpha$, in a.u.) for Zn.
            Error bars indicate the uncertainties from the numerical fitting procedure
            ($\Delta P_\text{fitting}$) for values with errors exceeding 0.005 a.u.
            }

\label{tab:dipole_group__12__Zn__2__0}
\begin{tabular}{llll}
\toprule
$\alpha$ (a.u.) & $\delta$ (\%) & Method & Comments \\
\midrule

$54.07 \pm 0.01$          & 33.60  & DHF     & 1C-NR-CC@s-aug-dyall.cv4z@(core 20)[vir 276]\\
$37.18 \pm 0.01 $         & -8.13  & MP2     &            \\
$41.99 \pm 0.01 $       & 3.76 & CCSD    &            \\
$40.47 \pm 0.01$  & $--$              & CCSD(T) &            \\
\cline{1-4}

$50.79 $          & 35.46  & DHF     & 1C-SR-CC@dyall.cv4z@(core 20)[vir 204]\\
$33.74  $         & -10.02  & MP2     &            \\
$38.98  $       & 3.96 & CCSD    &            \\
$37.49 $  & $--$              & CCSD(T) &            \\
\cline{1-4}

$50.84 $          & 33.84  & DHF     & 1C-SR-CC@s-aug-dyall.cv4z@(core 20)[vir 276]\\
$34.61  $         & -8.89  & MP2     &            \\
$39.43  $       & 3.79 & CCSD    &            \\
$37.99 $  & $--$              & CCSD(T) &            \\
\cline{1-4}

$50.84 $          & 33.84  & DHF     & 1C-SR-CC@d-aug-dyall.cv4z@(core 20)[vir 348]\\
$34.67  $         & -8.73  & MP2     &            \\
$39.43  $       & 3.79 & CCSD    &            \\
$37.99 $  & $--$              & CCSD(T) &            \\
\cline{1-4}
$50.81         $  & 33.51 & DHF     & 2C-DC-CC@s-aug-ANO-RCC@(core 20)[vir 314] \\
$34.69         $  & -8.85 & MP2     &             \\
$39.54       $  & 3.90 & CCSD    &             \\
$38.06 $  & $--$  & CCSD(T) &             \\
\cline{1-4}
$50.80         $  & 33.85 & DHF     & 2C-DC-CC@s-aug-dyall.cv4z@(core 20)[vir 276] \\
$34.57         $  & -8.90 & MP2     &             \\
$39.39       $  & 3.79 & CCSD    &             \\
$37.95 $  & $--$  & CCSD(T) &             \\

\bottomrule
\end{tabular}
\end{table}

Table~\ref{tab:dipole_group_12} summarizes the most accurate $\alpha$ values for Zn and other group 12 elements, compared with the recommended values from Ref.~\citenum{Schwerdtfeger2019}.
The corresponding $\gamma$ results are provided in Table S2 in the Supporting Information.
DC CCSD(T) values obtained using Eq.~\eqref{eq:ff_d_1} are also listed in Table~\ref{tab:dipole_group_12}, with additional data available in Table S3 in the Supporting Information.
A summary of the most accurate results from Eq.~\eqref{eq:ff_d_1} is provided in Table S4 in the Supporting Information.

\begin{table}
\centering
\caption{Static dipole polarizabilities (in a.u.) are presented with nonrelativistic, scalar-relativistic, and fully relativistic Dirac-Coulomb contributions for the elements of group 12. The uncertainty due to the numerical fitting procedure ($\Delta P_\text{fitting}$) is accounted for as the error bar. Only uncertainties where $\Delta P_\text{fitting} > 0.005$ a.u.\ are shown. The final recommended values (Rec.), along with their total uncertainty estimations shown as error bars, are provided and compared with the values reported in Ref.~\citenum{Schwerdtfeger2019}.}
\label{tab:dipole_group_12}
\scriptsize \begin{tabular}{llllll}
\toprule
                                 &      &                Zn &                Cd &                Hg &                Cn \\
$\hat{H}$ & Method &                   &                   &                   &                   \\
\midrule
NR & DHF &  $54.07 \pm 0.01$ &           $76.02$ &           $81.43$ &          $108.99$ \\
                                 & CCSD &  $41.99 \pm 0.01$ &           $57.12$ &           $60.19$ &           $79.52$ \\
                                 & CCSD(T) &  $40.47 \pm 0.01$ &           $54.24$ &  $56.87 \pm 0.01$ &  $74.02 \pm 0.01$ \\
SR & DHF &           $50.84$ &           $63.78$ &           $45.14$ &           $30.24$ \\
                                 & CCSD &           $39.43$ &           $48.08$ &           $35.43$ &           $27.78$ \\
                                 & CCSD(T) &           $37.99$ &           $45.76$ &           $34.21$ &  $27.75 \pm 0.01$ \\
DC & DHF &           $50.80$ &           $63.67$ &           $44.88$ &           $30.46$ \\
                                 & CCSD &           $39.39$ &           $48.00$ &           $35.25$ &           $27.96$ \\
                                 & CCSD(T) &           $37.95$ &           $45.68$ &           $34.04$ &  $27.94 \pm 0.02$ \\
DC (from Eq.~\eqref{eq:ff_d_1}) & CCSD(T) &           $38.00$ &           $45.75$ &           $34.07$ &           $27.93$ \\
DC (from Eq.~\eqref{eq:ff_d_1_fixed}) & CCSD(T) &              $--$ &              $--$ &              $--$ &           $27.92$ \\
Rec. & $--$ &  $37.95 \pm 0.77$ &  $45.68 \pm 1.21$ &  $34.04 \pm 0.68$ &  $27.92 \pm 0.28$ \\
Ref.~\citenum{Schwerdtfeger2019} & $--$ &     $38.67\pm0.3$ &          $46\pm2$ &    $33.91\pm0.34$ &          $28\pm2$ \\
\bottomrule
\end{tabular}
\end{table}

For Zn, the DC CCSD(T) $\alpha$ value obtained using Eq.~\eqref{eq:ff_d_2} is slightly lower than that obtained with Eq.~\eqref{eq:ff_d_1}, as shown in Table~\ref{tab:dipole_group_12}.
This suggests that applying Eq.~\eqref{eq:ff_d_2} provides a reasonable estimate of $\gamma$.

Theoretical and experimental values for Zn are summarized in Refs.~\citenum{Schwerdtfeger2019, Schwerdtfeger2023}, which also include detailed definitions and comments.
This work explicitly references the PRCC and NRCC methods.
The PRCC method with single and double excitations is denoted as PRCCSD, while PRCCSD with perturbative triple excitations is denoted as PRCCSD(T), PRCCSD$_p$T, or PRCC(T).
The NRCC method with single and double excitations is denoted as NRCCSD.

For consistency, the values are compiled in Table~\ref{tab:group_12_others_Zn}, organized by publication year, to validate the results of this study.
Only computational and experimental values close to the recommended values from Ref.~\citenum{Schwerdtfeger2019} are included.
For computational results, only values obtained using CC or its approximations are compared.

For Zn, the DC polarizability value (37.95 a.u.) is slightly below the lower bound of the recommended value ($38.67 \pm 0.30$ a.u.).\cite{Schwerdtfeger2019}
This recommended value aligns closely with the computational result obtained by Singh \textit{et al.} using PRCCSD$_p$T.\cite{Singh2014}
Our DC value using the ($31s21p13d7f5g3h$) basis is also lower than the value (38.75 a.u.) obtained using PRCC(T) with the ($17s15p15d10f10g9h$) basis\cite{Chattopadhyay2015} and the value ($38.99 \pm 0.31$ a.u.) computed using the RNCCSD method.\cite{Chakraborty2022}
Additionally, our DC value is slightly below the lower bound of the experimental result ($38.8 \pm 0.8$ a.u.).\cite{Goebel1996}
These differences may be attributed to the basis set employed in this work.

We also test the s-aug-ANO-RCC basis set ($22s16p11d7f5g3h$) at the same correlation level as s-aug-dyall.cv4z.
The DC value obtained using Eq.~\eqref{eq:ff_d_2} for s-aug-ANO-RCC is 38.06 a.u., as shown in Table~\ref{tab:dipole_group_12}, which falls within the experimental uncertainty.\cite{Goebel1996}
Given the minor differences between the s-aug-ANO-RCC and s-aug-dyall.cv4z results, we recommend the DC value computed with s-aug-dyall.cv4z as the final value due to its larger basis set.
Our SR result (37.99 a.u.) closely matches the CCSD(T) value (37.7 a.u.) obtained using the effective core potential (ECP) method.\cite{Zaremba-Kopczyk2021}

\begin{table}[ht]
\centering
\caption{
    Summary of reference atomic dipole polarizabilities (in a.u.) for the ground state of Zn, as reported in
    Refs.\citenum{Schwerdtfeger2019} and \citenum{Schwerdtfeger2023} where the definitions of comments are available therein.
    Data from Ref.~\citenum{Schwerdtfeger2019} are reprinted with permission. Copyright 2018 by Taylor \& Francis.
    Data from Ref.~\citenum{Schwerdtfeger2023} are reprinted with permission. Copyright 2023 by Peter Schwerdtfeger and Jeffrey K. Nagle.
    }
\label{tab:group_12_others_Zn}
\begin{tabular}{lllrl}
\toprule
&  Refs. & $\alpha$ & Year & Comments \\
\midrule
& [\citenum{Kellö1995}]                   & $37.6$ & 1995 & R, MVD, CCSD(T) \\
& [\citenum{Goebel1996}]                  & $39.2 \pm 0.8$ & 1996 & NR, CCSD(T), MP2 basis correction \\
& [\citenum{Goebel1996}]                  & $38.8 \pm 0.8$ & 1996 & exp. \\
& [\citenum{Seth1997}]                    & $38.01$ & 1997 & R, PP, CCSD(T) \\
& [\citenum{Ellingsen2001}]               & $39.12$ & 2001 & R, MRCI, pseudo-potential \\
& [\citenum{Lide2004, Doolen1987}]        & $38 \pm 9$ & 2004 & R, Dirac, LDA \\
& [\citenum{Chu2004}]                     & $37.7$ & 2004 & SIC-DFT \\
& [\citenum{Roos2005}]                    & $38.4$ & 2005 & R, DK, CASPT2 \\
& [\citenum{Maroulis2006, Kellö1995}]     & $38.35 \pm 0.29$ & 2006 & R, MVD, CCSD(T) \\
& [\citenum{Singh2014}]                   & $38.666 \pm 0.096$ & 2014 & R, Dirac, PRCCSD$_p$T \\
& [\citenum{Chattopadhyay2015}]           & $38.75$ & 2015 & R, PRCC(T) \\
& [\citenum{Chattopadhyay2015, Qiao2012}] & $38.92$ & 2015 & exp.+fitting \\
& [\citenum{Gould2016b}]                  & $39.2$ & 2016 & SIC-DFT (RXH) \\
& [\citenum{Szarek2019}]                  & $41.50$ & 2019 & R, CCSD(T)/ANO-RCC \\
& [\citenum{Schwerdtfeger2019}]           & $38.67 \pm 0.30$ & 2019 & recommended \\
& [\citenum{Zaremba-Kopczyk2021}]         & $37.7$ & 2021 & ECP, CCSD(T) \\
& [\citenum{Chakraborty2022}]             & $38.99 \pm 0.31$ & 2022 & R, Dirac, RNCCSD \\
\bottomrule
\end{tabular}
\end{table}

\subsection{Cd}

Table~\ref{tab:dipole_group__12__Cd__2__0} presents the dipole polarizabilities $\alpha$ of Cd, determined by fitting Eq.~\eqref{eq:ff_d_2}.
The corresponding hyperpolarizabilities $\gamma$ are listed in Table S1 in the Supporting Information.
As with Zn, calculations using the s-aug-dyall.cv4z basis set are adopted as the most accurate method at each relativistic level.

\begin{table}[ht]
\centering

            \caption{
            Same as Table~\ref{tab:dipole_group__12__Zn__2__0} but for Cd.
            }

\label{tab:dipole_group__12__Cd__2__0}
\begin{tabular}{llll}
\toprule
$\alpha$ (a.u.) & $\delta$ (\%) & Method & Comments \\
\midrule

$76.02 $          & 40.14  & DHF     & 1C-NR-CC@s-aug-dyall.cv4z@(core 30)[vir 344]\\
$44.80  $         & -17.41  & MP2     &            \\
$57.12  $       & 5.31 & CCSD    &            \\
$54.24 $  & $--$              & CCSD(T) &            \\
\cline{1-4}

$63.76 $          & 39.90  & DHF     & 1C-SR-CC@dyall.cv4z@(core 30)[vir 272]\\
$36.65  $         & -19.58  & MP2     &            \\
$47.92  $       & 5.16 & CCSD    &            \\
$45.57 $  & $--$              & CCSD(T) &            \\
\cline{1-4}

$63.78 $          & 39.38  & DHF     & 1C-SR-CC@s-aug-dyall.cv4z@(core 30)[vir 344]\\
$37.19  $         & -18.72  & MP2     &            \\
$48.08  $       & 5.08 & CCSD    &            \\
$45.76 $  & $--$              & CCSD(T) &            \\
\cline{1-4}

$63.78 $          & 39.37  & DHF     & 1C-SR-CC@d-aug-dyall.cv4z@(core 30)[vir 416]\\
$37.21  $         & -18.69  & MP2     &            \\
$48.09  $       & 5.08 & CCSD    &            \\
$45.76 $  & $--$              & CCSD(T) &            \\
\cline{1-4}
$63.59         $  & 39.59 & DHF     & 2C-DC-CC@s-aug-ANO-RCC@(core 30)[vir 366] \\
$38.08         $  & -16.40 & MP2     &             \\
$47.59       $  & 4.47 & CCSD    &             \\
$45.55 $  & $--$  & CCSD(T) &             \\
\cline{1-4}
$63.67         $  & 39.39 & DHF     & 2C-DC-CC@s-aug-dyall.cv4z@(core 30)[vir 344] \\
$37.12         $  & -18.72 & MP2     &             \\
$48.00       $  & 5.08 & CCSD    &             \\
$45.68 $  & $--$  & CCSD(T) &             \\

\bottomrule
\end{tabular}
\end{table}

The most accurate $\alpha$ values for Cd, along with the recommended values from Ref.~\citenum{Schwerdtfeger2019}, are summarized in Table~\ref{tab:dipole_group_12}.
The DC CCSD(T) results obtained using Eq.~\eqref{eq:ff_d_1} are also included, with additional data provided in Table S3 in the Supporting Information.

The DC value (45.68 a.u.) aligns closely with the recommended value ($46 \pm 2$ a.u.) from Ref.~\citenum{Schwerdtfeger2019}.
Guo \textit{et al.} used a similar procedure with the d-aug-dyall.v4z basis set, obtaining dipole and hyperpolarizability values of $46.005$ a.u.\ and $40570$ a.u., respectively.\cite{Guo2021}
Their final recommended values were $45.92$ a.u.\ and $40628$ a.u., respectively, showing differences of less than 0.71\% and 4.9\% for $\alpha$ and $\gamma$, respectively, compared to this work.
The slight discrepancies in $\alpha$ may arise from differences in the number of correlated electrons and energy cutoffs for virtual spinors used in the two studies.
The DHF, CCSD, and CCSD(T) energies with the s-aug-dyall.cv4z basis set in this work are -5593.442, -5594.510, and -5594.541 a.u., respectively, slightly lower than the corresponding values of -5593.318, -5594.312, and -5594.342 a.u.\ obtained with the d-aug-dyall.v4z basis.\cite{Guo2021}
Additionally, this work employs a least-squares procedure with five data points, compared to the two-point numerical differentiation used in Ref.~\citenum{Guo2021}.

Our DC results also agree well with computational values of $45.86 \pm 0.15$ a.u.\ obtained using PRCCSD$_p$T,\cite{Singh2014} and $46.02 \pm 0.50$ a.u.\ recommended by Sahoo and Yu using a set of CC methods.\cite{Sahoo2018b}
The current DC values are consistent with experimental measurements ($45.3 \pm 1.4$ \cite{Bromley2002a, Goebel1995a} and $47.5 \pm 2.0$ a.u.\cite{Hohm2022a}), but they are slightly below the lower bounds of other experimental results ($49.7 \pm 1.6$ \cite{Goebel1995} and $48.2 \pm 1.1$ a.u.\cite{Goebel1995a}).

\begin{table}[ht]
\centering
\caption{
    Same as Table~\ref{tab:group_12_others_Zn} but for Cd.
}
\label{tab:group_12_others_Cd}
\begin{tabular}{lllrl}
\toprule
&  Refs.  & $\alpha$ & Year & Comments \\
\midrule
& [\citenum{Kellö1995}]                 & $46.8$           & 1995 & R, MVD, CCSD(T) \\
& [\citenum{Goebel1995}]                & $49.7 \pm 1.6$   & 1995 & exp. \\
& [\citenum{Goebel1995a}]               & $48.2 \pm 1.1$   & 1995 & exp. \\
& [\citenum{Seth1997}]                  & $46.25$          & 1997 & R, PP, CCSD(T) \\
& [\citenum{Bromley2002a, Goebel1995a}] & $45.3 \pm 1.4$   & 2002 & exp. \\
& [\citenum{Moszynski2003}]             & $45.91/53.99$    & 2003 & CCSD R/NR \\
& [\citenum{Roos2005}]                  & $46.9$           & 2005 & R, DK, CASPT2 \\
& [\citenum{Maroulis2006, Kellö1995}]   & $47.55 \pm 0.48$ & 2006 & R, MVD, CCSD(T) \\
& [\citenum{Ye2008}]                    & $44.63$          & 2008 & R, DHF, CPMP \\
& [\citenum{Singh2014}]                 & $45.86 \pm 0.15$ & 2014 & R, DF, PRCCSD$_p$T, MBPT3 \\
& [\citenum{Gould2016a}]                & $46.7$           & 2016 & TD-DFT (LEXX) \\
& [\citenum{Sahoo2018b}]                & $46.02 \pm 0.50$ & 2018 & R, Dirac, CCSD(T) \\
& [\citenum{A.Manz2019}]                & $48.3$           & 2019 & ECP, CCSD \\
& [\citenum{Schwerdtfeger2019}]         & $46 \pm 2$       & 2019 & recommended \\
& [\citenum{Dutta2020}]                 & $39.79$          & 2020 & R, Dirac, MBPT3 \\
& [\citenum{Guo2021}]                   & $45.92 \pm 0.10$ & 2021 & R, Dirac, CCSD(T) \\
& [\citenum{Zaremba-Kopczyk2021}]       & $45.8$           & 2021 & ECP, CCSD(T) \\
& [\citenum{Zhou2021a}]                 & $46 \pm 2$       & 2021 & R, DFCP+RCI \\
& [\citenum{Hohm2022a}]                 & $47.5 \pm 2.0$   & 2022 & exp. \\
\bottomrule
\end{tabular}
\end{table}

\subsection{Hg}

Table~\ref{tab:dipole_group__12__Hg__2__0} presents the dipole polarizabilities $\alpha$ of Hg, determined by fitting Eq.~\eqref{eq:ff_d_2}.
The corresponding hyperpolarizabilities $\gamma$ are listed in Table S1 in the Supporting Information.
As with Zn and Cd, calculations using the s-aug-dyall.cv4z basis set are adopted as the most accurate method at each relativistic level.

\begin{table}[ht]
\centering

            \caption{
            Same as Table~\ref{tab:dipole_group__12__Zn__2__0} but for Hg.
            }

\label{tab:dipole_group__12__Hg__2__0}
\begin{tabular}{llll}
\toprule
$\alpha$ (a.u.) & $\delta$ (\%) & Method & Comments \\
\midrule

$81.43 $          & 43.18  & DHF     & 1C-NR-CC@s-aug-dyall.cv4z@(core 44)[vir 362]\\
$44.17  $         & -22.33  & MP2     &            \\
$60.19 \pm 0.01 $       & 5.84 & CCSD    &            \\
$56.87 \pm 0.01$  & $--$              & CCSD(T) &            \\
\cline{1-4}

$45.08 $          & 32.30  & DHF     & 1C-SR-CC@dyall.cv4z@(core 44)[vir 290]\\
$27.20  $         & -20.18  & MP2     &            \\
$35.32  $       & 3.64 & CCSD    &            \\
$34.08 $  & $--$              & CCSD(T) &            \\
\cline{1-4}

$45.14 $          & 31.94  & DHF     & 1C-SR-CC@s-aug-dyall.cv4z@(core 44)[vir 362]\\
$27.46  $         & -19.72  & MP2     &            \\
$35.43  $       & 3.56 & CCSD    &            \\
$34.21 $  & $--$              & CCSD(T) &            \\
\cline{1-4}

$45.14 $          & 32.21  & DHF     & 1C-SR-CC@d-aug-dyall.cv4z@(core 44)[vir 460]\\
$27.35  $         & -19.88  & MP2     &            \\
$35.37 \pm 0.01 $       & 3.61 & CCSD    &            \\
$34.14 \pm 0.01$  & $--$              & CCSD(T) &            \\
\cline{1-4}
$44.82         $  & 30.50 & DHF     & 2C-DC-CC@s-aug-ANO-RCC@(core 44)[vir 282] \\
$28.04         $  & -18.34 & MP2     &             \\
$35.47    \pm 0.01   $  & 3.29 & CCSD    &             \\
$34.34 \pm 0.01$  & $--$  & CCSD(T) &             \\
\cline{1-4}
$44.88         $  & 31.86 & DHF     & 2C-DC-CC@s-aug-dyall.cv4z@(core 44)[vir 362] \\
$27.37         $  & -19.59 & MP2     &             \\
$35.25       $  & 3.56 & CCSD    &             \\
$34.04 $  & $--$  & CCSD(T) &             \\

\bottomrule
\end{tabular}
\end{table}

The most accurate $\alpha$ values for Hg, along with the recommended values from Ref.~\citenum{Schwerdtfeger2019}, are summarized in Table~\ref{tab:dipole_group_12}.
The DC CCSD(T) results obtained using Eq.~\eqref{eq:ff_d_1} are also included, with additional data provided in Table S3 in the Supporting Information.

For Hg, the DC polarizability value (34.04 a.u.) agrees well with the recommended value ($33.91 \pm 0.34$ a.u.) from Ref.~\citenum{Schwerdtfeger2019}, which is based on experimental results from Ref.~\citenum{Goebel1996a}.
Our DC value also aligns with other CCSD(T) values reported in the literature, including $34.15$ a.u.\cite{Pershina2008b} and $34.1$ a.u.\cite{Borschevsky2015}
The DC basis set used in this work is ($35s31p20d14f8g5h2i$), compared to ($26s24p18d13f5g2h$) in Refs.~\citenum{Pershina2008b, Borschevsky2015}.
Additionally, our DC value closely matches results obtained using the RNCCSD and PRCCSD(T) methods, such as $34.2 \pm 0.5$ a.u.\cite{Sahoo2018d} and $34.5 \pm 0.8$ a.u.,\cite{Sahoo2018} respectively.
The final recommended $\alpha$ in Ref.~\citenum{Sahoo2018d} includes a Breit contribution of $-0.01$ a.u.
However, the triple contribution to $\alpha$ differs, with $-0.28$ a.u.\ in Ref.~\citenum{Sahoo2018d} and $-1.21$ a.u.\ in this work, as shown in Table~\ref{tab:dipole_group_12}.

Our DC value is also comparable to PRCCSD$_p$T values of 34.07 and 34.27 a.u., calculated without and with Breit and QED corrections, respectively, in Ref.~\citenum{Singh2015}.
It is further consistent with the PRCC(T) value ($33.69 \pm 0.34$ a.u.) reported by Kumar \textit{et al.}, which includes Breit and QED corrections.\cite{Kumar2021}

Our SR value (34.21 a.u.) aligns well with previous results, such as $34.73 \pm 0.52$ a.u.\ obtained using the DK CCSD(T) method.\cite{Maroulis2006, Kellö1996}
Additionally, the SR CCSD value is in good agreement with the result (35.45 a.u.) obtained using the CCSD method with the ECP approach.\cite{A.Manz2019}

\begin{table}[ht]
\centering
\caption{
    Same as Table~\ref{tab:group_12_others_Zn} but for Hg.
    }
\label{tab:group_12_others_Hg}
\begin{tabular}{lllrl}
\toprule
& Refs. & $\alpha$ & Year & Comments \\
\midrule
& [\citenum{Kellö1995}]                     & $31.24$          & 1995 & R, MVD, CCSD(T) \\
& [\citenum{Goebel1996a}]                   & $33.91 \pm 0.34$ & 1996 & exp. \\
& [\citenum{Seth1997}]                      & $34.42$          & 1997 & R, PP, CCSD(T) \\
& [\citenum{Roos2005}]                      & $33.3$           & 2005 & R, DK, CASPT2 \\
& [\citenum{Maroulis2006, Kellö1996}]       & $34.73 \pm 0.52$ & 2006 & R, DK, CCSD(T) \\
& [\citenum{Pershina2008b}]                 & $34.15$          & 2008 & R, Dirac, CCSD(T) \\
& [\citenum{Kellö1995, Qiao2012, Tang2008}] & $33.75$          & 2012 & exp. \\
& [\citenum{Singh2015}]                     & $34.27$          & 2015 & R, Dirac, PRCCSD$_p$T + Breit + QED \\
& [\citenum{Borschevsky2015}]               & $34.1$           & 2015 & R, Dirac, CCSD(T) \\
& [\citenum{Chattopadhyay2015}]             & $33.59$          & 2015 & R, Dirac, PRCC(T) \\
& [\citenum{Dyugaev2016}]                   & $32.9$           & 2016 & semi-empirical \\
& [\citenum{Dzuba2016b}]                    & $39.1$           & 2016 & R, RPA, PolPot \\
& [\citenum{gobre2016efficient}]            & $33.90$          & 2016 & LR-CCSD \\
& [\citenum{Gould2016a}]                    & $33.5$           & 2016 & TD-DFT (LEXX) \\
& [\citenum{Sahoo2018d}]                    & $34.2 \pm 0.5$   & 2018 & R, Dirac, RNCCSD + Triples + Breit + Basis \\
& [\citenum{Sahoo2018}]                     & $34.5 \pm 0.8$   & 2018 & R, Dirac, PRCCSD(T) \\
& [\citenum{A.Manz2019}]                    & $35.45$          & 2019 & ECP, CCSD \\
& [\citenum{Schwerdtfeger2019}]             & $33.91 \pm 0.34$ & 2019 & recommended \\
& [\citenum{Kumar2021}]                     & $33.69 \pm 0.34$ & 2021 & R, Dirac, PRCC(T) + Breitt + QED \\
& [\citenum{Centoducatte2022}]              & $36.1$           & 2022 & R (ZORA), DFT (B3LYP) \\
& [\citenum{Neto2023}]                      & $34.9$           & 2023 & R (ATZP-ZORA), DFT (B3LYP) \\
\bottomrule
\end{tabular}
\end{table}

\subsection{Cn}

Table~\ref{tab:dipole_group__12__Cn__2__0} presents the dipole polarizabilities $\alpha$ of Cn, determined by fitting Eq.~\eqref{eq:ff_d_2}.
The corresponding hyperpolarizabilities $\gamma$ are provided in Table S1 in the Supporting Information.
As with Zn, Cd, and Hg, calculations using the s-aug-dyall.cv4z basis set are adopted as the most accurate method at each relativistic level.

\begin{table}[ht]
\centering

            \caption{
            Same as Table~\ref{tab:dipole_group__12__Zn__2__0} but for Cn.
            }

\label{tab:dipole_group__12__Cn__2__0}
\begin{tabular}{llll}
\toprule
$\alpha$ (a.u.) & $\delta$ (\%) & Method & Comments \\
\midrule

$108.99 $          & 47.24  & DHF     & 1C-NR-CC@s-aug-dyall.cv4z@(core 44)[vir 434]\\
$52.65  $         & -28.87  & MP2     &            \\
$79.52 \pm 0.01 $       & 7.43 & CCSD    &            \\
$74.02 \pm 0.01$  & $--$              & CCSD(T) &            \\
\cline{1-4}

$30.24 $          & 8.78  & DHF     & 1C-SR-CC@dyall.cv4z@(core 44)[vir 336]\\
$25.82  $         & -7.13  & MP2     &            \\
$27.83 \pm 0.01 $       & 0.09 & CCSD    &            \\
$27.80 \pm 0.01$  & $--$              & CCSD(T) &            \\
\cline{1-4}

$30.24 $          & 8.99  & DHF     & 1C-SR-CC@s-aug-dyall.cv4z@(core 44)[vir 434]\\
$25.78  $         & -7.09  & MP2     &            \\
$27.78 \pm 0.01 $       & 0.11 & CCSD    &            \\
$27.75 \pm 0.01$  & $--$              & CCSD(T) &            \\
\cline{1-4}

$30.25 $          & 9.04  & DHF     & 1C-SR-CC@d-aug-dyall.cv4z@(core 44)[vir 532]\\
$25.75  $         & -7.17  & MP2     &            \\
$27.77 \pm 0.01 $       & 0.12 & CCSD    &            \\
$27.74 \pm 0.01$  & $--$              & CCSD(T) &            \\
\cline{1-4}
$30.51         $  & 8.71 & DHF     & 4C-DC-CC@dyall.cv4z@(core 12)[vir 264] \\
$26.92         $  & -4.07 & MP2     &             \\
$27.97       $  & -0.33 & CCSD    &             \\
$28.07 $  & $--$  & CCSD(T) &             \\
\cline{1-4}
$30.46         $  & 9.01 & DHF     & 2C-DC-CC@s-aug-dyall.cv4z@(core 48)[vir 434] \\
$25.97         $  & -7.08 & MP2     &             \\
$27.96    \pm 0.02   $  & 0.06 & CCSD    &             \\
$27.94 \pm 0.02$  & $--$  & CCSD(T) &             \\

\bottomrule
\end{tabular}
\end{table}

The most accurate $\alpha$ values for Cn, along with the recommended values from Ref.~\citenum{Schwerdtfeger2019}, are summarized in Table~\ref{tab:dipole_group_12}.
The DC CCSD(T) results obtained using Eq.~\eqref{eq:ff_d_1} are also included, with additional data provided in Table S3 in the Supporting Information.

For Cn, the central DC CCSD(T) $\alpha$ value obtained using Eq.~\eqref{eq:ff_d_2} is slightly higher than the result from Eq.~\eqref{eq:ff_d_1}, as shown in Table~\ref{tab:dipole_group_12}.
This discrepancy arises because $\gamma$, defined in Eq.~\eqref{eq:ff_d_2}, is negative for Cn when DC CCSD(T) energies are used (Table S2, Supporting Information).
Since $\gamma$ should ideally be positive, this suggests a potential overestimation of the DC $\alpha$ values.
However, both SR and DC hyperpolarizabilities are negative for calculations with 30 core electrons, as shown in Table S2 in the Supporting Information.
Given the small difference between the DC $\alpha$ obtained by including only 12 core electrons with the dyall.cv4z basis set and 30 core electrons with s-aug-dyall.cv4z, $\gamma= 0.37 \times 10^4$ is taken as the approximate $\gamma$ in this work.
The DC value obtained using Eq.~\eqref{eq:ff_d_1_fixed} is $27.92$ a.u., which closely matches the DC values obtained from Eq.~\eqref{eq:ff_d_1} ($27.93$ a.u.) and Eq.~\eqref{eq:ff_d_2} ($27.94 \pm 0.02$ a.u.) due to the small magnitude of $\gamma$.
For simplicity, $27.92$ a.u.\ is used as the final recommended DC value.
The DC CCSD(T) value ($27.94 \pm 0.02$ a.u.) obtained using Eq.~\eqref{eq:ff_d_2} is used in subsequent discussions, as the small difference (0.02 a.u.) between results from Eqs.~\eqref{eq:ff_d_2} and \eqref{eq:ff_d_1} is negligible, despite a negative $\gamma$ from Eq.~\eqref{eq:ff_d_2}.

Our DC value ($27.94 \pm 0.02$ a.u.) is consistent with the recommended value ($28 \pm 2$ a.u.) from Ref.~\citenum{Schwerdtfeger2019}.
Pershina \textit{et al.} computed $\alpha$ of Cn using relativistic CCSD(T) with a smaller basis set ($26s24p18d13f5g2h$) and a numerical differentiation method with three data points, correlating 36 electrons.
Their final values, $27.64$ and $27.40$ a.u., include and exclude a correction of 0.24 a.u.\ (based on the difference between the experimental value and DC value for Hg), respectively.\cite{Pershina2008b}
In this work, the s-aug-dyall.cv4z basis ($37s36p25d17f7g5h2i$) is used, with 48 electrons correlated and a least-squares procedure applied using five data points.
Our DC value aligns closely with the recent computational result ($27.44 \pm 0.88$ a.u.) obtained using PRCC(T) with Breit and QED corrections.\cite{Kumar2021}

The DC CCSD, SR CCSD, and SR CCSD(T) values in this work are 27.96, 27.78, and $27.75 \pm 0.01$ a.u., respectively, indicating a minimal impact of SOC and higher-order correlation effects on $\alpha$ when SR effects are included.
The triple contribution to $\alpha$ at the DC level is -0.07\%, consistent with previous findings of -0.08\%\cite{Seth1997} and -0.07\%.\cite{Pershina2008b}

\begin{table}[ht]
\centering
\caption{
    Same as Table~\ref{tab:group_12_others_Zn} but for Cn.
    }
\label{tab:group_12_others_Cn}
\begin{tabular}{lllrl}
\toprule
&  Refs. & $\alpha$ & Year & Comments \\
\midrule
& [\citenum{Seth1997}]          & $25.82$ & 1997 & R, PP, CCSD(T) \\
& [\citenum{Nash2005}]          & $28.68$ & 2005 & R, SOPP, CCSD(T) \\
& [\citenum{Pershina2008b}]     & $27.64$ & 2008 & R, Dirac, CCSD(T) \\
& [\citenum{Pershina2008b}]     & $27.40$ & 2008 & R, Dirac, CCSD(T) \\
& [\citenum{Dzuba2016b}]        & $28.2$  & 2016 & R, RPA, PolPot \\
& [\citenum{Dzuba2016b}]        & $28 \pm 4$ & 2016 & R, RPA, PolPot (value recommended by authors) \\
& [\citenum{Schwerdtfeger2019}] & $28 \pm 2$ & 2019 & recommended \\
& [\citenum{Kumar2021}]         & $27.44 \pm 0.88$ & 2021 & R, Dirac, PRCC(T) + Breit + QED \\
\bottomrule
\end{tabular}
\end{table}

\subsection{Uncertainty estimation}
\label{ssec:reliable-group-12}

Next, the uncertainty excluding $\Delta P_\text{fitting}$ is estimated for all group 12 elements.
Half of the difference between SR values evaluated by s-aug-dyall.cv4z and d-aug-dyall.cv4z is taken as the error due to the finite basis set, $\Delta P_\text{basis}$ for each atom.
Half of the difference between DC CCSD and CCSD(T) results is used as the error due to $\Delta P_\text{(T)}$.
The corresponding errors are 0.72, 1.16, 0.60, and 0.01 a.u. for Zn, Cd, Hg, and Cn, respectively.
The SOC effect obtained by Eq.~\eqref{eq:ff_d_2} is used as the error due to $\Delta P_\text{SOC}$.

Dutta \textit{et al.} studied the Gaunt or Breit effects on $\alpha$ for group 12 elements, excluding Cn, using perturbation theory.
The corresponding values calculated using third-order perturbation theory are 0.033, -0.0385, and -0.2233 a.u. for Zn, Cd, and Hg, respectively.
These results suggest that the magnitude of the Gaunt or Breit contributions to $\alpha$ increases with $Z$ for Zn, Cd, and Hg.
Kumar \textit{et al.} investigated the Breit and QED contributions for all group 12 elements using the PRCC method.\cite{Kumar2021}
However, the magnitude of the Breit contribution decreases with increasing $Z$, except for Cn.
In addition, the QED contributions (including vacuum polarization and self-energy corrections) to the DC $\alpha$ increase with $Z$ for group 12 elements.
Based on Ref.~\citenum{Kumar2021}, the upper bound of the frequency-dependent Breit interaction is approximately 0.13\% of the DC $\alpha$ for Cn, while the higher-order QED contributions are less than 0.1\% of the DC $\alpha$ for Cn.
Therefore, the combined upper bound of the Breit and QED contributions (0.5\%) to the DC $\alpha$ of Cn is approximately 0.73\%.
To remain conservative, we adopt the highest contribution of 0.73\% for all group 12 elements to account for the uncertainty associated with $\Delta P_\text{others}$.

In conclusion, the total errors, excluding $\Delta P_\text{fitting}$, are 0.77, 1.21, 0.67, and 0.28 a.u. for Zn, Cd, Hg, and Cn, respectively.
The final recommended values (Rec.), obtained from the most accurate calculations with total uncertainty, including the corresponding $\Delta P_\text{fitting}$, are $\recZn$, $\recCd$, $\recHg$, and $\recCn$ a.u. for Zn, Cd, Hg, and Cn, respectively, as listed in \autoref{tab:dipole_group_12}.
The total error for Cd is the largest compared to the other group 12 elements, while Cn has the smallest total error.
This is due to the conservative error estimation for high-order electron correlation, which is the dominant source of error for all group 12 elements in this work.

\subsection{The correlation and relativistic effects on polarizabilities}

\begin{figure}[h]
    \centering
    \includegraphics[scale=1]{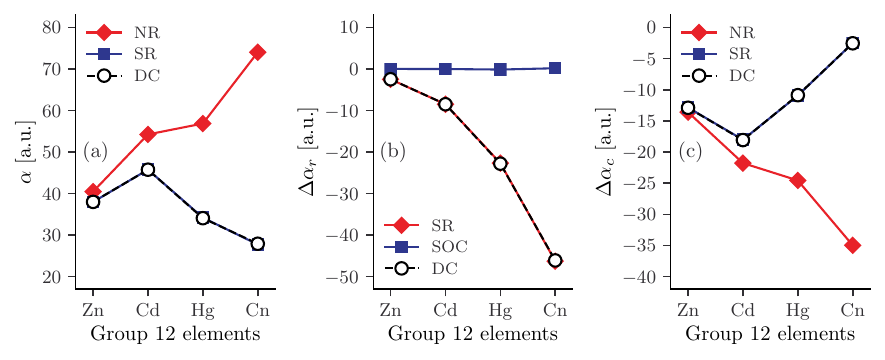}
    \caption{
    Dipole polarizabilities (in a.u.) of group 12 elements.
    (a) Dependence of NR, SR, and DC dipole polarizabilities on atomic number.
    (b) Contribution of SR, SOC, and DC relativistic effects on dipole polarizabilities.
    (c) Contribution of electron correlation to dipole polarizabilities in the presence of NR, SR, and DC effects.
    }
    \label{fig:group_12}
\end{figure}

The relationship between dipole polarizabilities and atomic numbers for group 12 elements is depicted in Fig.~\ref{fig:group_12}(a).
Nonrelativistic polarizabilities of group 12 elements increase with atomic number.
Considering relativistic effects, Cd has the largest values, 45.76 and 45.68 for both scalar-relativistic and DC values, respectively.
Relativistic values decrease with increasing atomic number starting from Cd.
As a result, Zn has a larger DC value than Hg and Cn.
This irregular trend is attributed to the relativistic contraction of the valence $s$-shell, analogous to that observed in $s$-block elements.\cite{Cheng2024a}
The trend observed in the DC results is consistent with that of scalar-relativistic calculations, indicating a very small SOC effect contribution for group 12 elements.
This observation is further supported by Fig.~\ref{fig:group_12}(b), where SR and DC data perfectly overlap.

The trend in electron correlation is presented in Fig.~\ref{fig:group_12}(c).
At the nonrelativistic level, electron correlation contributions increase with atomic number.
In relativistic calculations starting from Cd, electron correlation contributions decrease with increasing atomic number.

    \section{Summary}
    \label{sec:summary}
    In this work, we calculated the static dipole polarizabilities of group 12 elements using the finite-field method combined with relativistic CCSD(T) calculations.
The recommended polarizability values, with their associated uncertainties, are $\recZn$ a.u.\ for Zn, $\recCd$ a.u.\ for Cd, $\recHg$ a.u.\ for Hg, and $\recCn$ a.u.\ for Cn, showing excellent agreement with previously reported values in the literature.
Moreover, a systematic analysis was carried out to separate the contributions of scalar-relativistic effects, SOC, and fully relativistic Dirac-Coulomb effects.
Our findings show that scalar-relativistic effects dominate the relativistic corrections for group 12 elements, while SOC effects are negligible.
The impact of electron correlation, combined with different relativistic corrections, on atomic dipole polarizabilities was also investigated.
The results demonstrate that electron correlation plays a critical role in achieving accurate polarizability calculations.

    \begin{acknowledgement}
            Y.C. acknowledges the Foundation of Scientific Research - Flanders (FWO, file number G0A9717N) and the Research Board of Ghent University (BOF) for their financial support.
    The resources and services used in this work were provided by the VSC (Flemish Supercomputer Center), funded by the Research Foundation - Flanders (FWO) and the Flemish Government.
     \end{acknowledgement}

    \begin{suppinfo}
        The Supplementary Material provides a PDF document with the atomic dipole hyperpolarizabilities, obtained by fitting Eq.~\eqref{eq:ff_d_2}, and the dipole polarizabilities, obtained by fitting Eq.~\eqref{eq:ff_d_1}, for group 12 elements.
     \end{suppinfo}

    \providecommand{\latin}[1]{#1}
\makeatletter
\providecommand{\doi}
  {\begingroup\let\do\@makeother\dospecials
  \catcode`\{=1 \catcode`\}=2 \doi@aux}
\providecommand{\doi@aux}[1]{\endgroup\texttt{#1}}
\makeatother
\providecommand*\mcitethebibliography{\thebibliography}
\csname @ifundefined\endcsname{endmcitethebibliography}
  {\let\endmcitethebibliography\endthebibliography}{}

\end{document}